\documentclass[
reprint,
longbibliography,
superscriptaddress,
preprintnumbers,
nobibnotes,
amsmath,amssymb,
aps,
pra,
floatfix
]{revtex4-2}

\usepackage{graphicx}
\usepackage{dcolumn}
\usepackage{bm}
\usepackage{floatrow}
\usepackage[%
  colorlinks=true,
  urlcolor=blue,
  linkcolor=blue,
  citecolor=blue
]{hyperref}
\usepackage{xcolor}
\usepackage{textcomp}
\usepackage{amsmath}
\usepackage[artemisia]{textgreek}
\usepackage{upgreek}
\usepackage{etoolbox}
\usepackage[version=4]{mhchem} 

\usepackage{siunitx}

\usepackage{multirow}

\usepackage{dcolumn}
\newcolumntype{z}[1]{D{.}{.}{#1}}

\usepackage{amsmath}
\usepackage{pbox}
\usepackage{braket}

\usepackage[super]{nth}

\begin{document}

\title{Temperature Sensitivity of $^{14}\mathrm{NV}$ and $^{15}\mathrm{NV}$ Ground State Manifolds}

    \author{Sean~Lourette}
    \email{slourette@berkeley.edu}
    \affiliation{
     Department of Physics, University of California,
     Berkeley, California 94720, USA
     }
      \affiliation{
    DEVCOM Army Research Laboratory, Adelphi, Maryland 20783, USA 
    }

    \author{Andrey~Jarmola}
    \affiliation{
     Department of Physics, University of California,
     Berkeley, California 94720, USA
     }
      \affiliation{
    DEVCOM Army Research Laboratory, Adelphi, Maryland 20783, USA 
    }    

    \author{Victor~M.~Acosta}
    \affiliation{
    Center for High Technology Materials and Department of Physics and Astronomy,
University of New Mexico, Albuquerque, New Mexico 87106, USA 
    }  
    
    \author{A.~Glen~Birdwell}
    \affiliation{
    DEVCOM Army Research Laboratory, Adelphi, Maryland 20783, USA 
    }   

    \author{Dmitry~Budker}
    \affiliation{
     Department of Physics, University of California,
     Berkeley, California 94720, USA
     }
    \affiliation{Johannes Gutenberg-Universit{\"a}t Mainz, 55128 Mainz, Germany}
    \affiliation{Helmholtz-Institut, GSI Helmholtzzentrum f{\"u}r Schwerionenforschung, 55128 Mainz, Germany}
    
    \author{Marcus~W.~Doherty}
    \affiliation{
    Department of Quantum Science \& Technology, Research School of Physics, Australian National University, Canberra 2601, Australia.
    }

    \author{Tony~Ivanov}
    \affiliation{
    DEVCOM Army Research Laboratory, Adelphi, Maryland 20783, USA 
    }

 \author{Vladimir~S.~Malinovsky}
    \affiliation{
    DEVCOM Army Research Laboratory, Adelphi, Maryland 20783, USA 
    }

\date{\today}

\begin{abstract}

We measure electron and nuclear spin transition frequencies in the ground state of nitrogen-vacancy (NV) centers in diamond 
for two nitrogen isotopes ($^{14}\mathrm{NV}$ and $^{15}\mathrm{NV}$)
over temperatures ranging from 77\,K to 400\,K.
Measurements are performed using Ramsey interferometry and direct optical readout of the nuclear and electron spins.
We extract coupling parameters $Q$ (for $^{14}\mathrm{NV}$), $D$, $A_{||}$, $A_{\perp}$, $\gamma_e/\gamma_n$, and their temperature dependences for both isotopes.
The temperature dependences of the nuclear-spin transitions within the $m_s = 0$ spin manifold
near room temperature
are found to be \SI{+0.52(1)}{ppm/K} for $^{14}\mathrm{NV}$ ($\ket{m_I=-1}\leftrightarrow \ket{m_I=+1}$) and \SI{-1.1(1)}{ppm/K} for $^{15}\mathrm{NV}$ ($\ket{m_I=-1/2}\leftrightarrow \ket{m_I=+1/2}$).
An isotopic shift in the zero-field splitting parameter $D$ between $^{14}\mathrm{NV}$ and $^{15}\mathrm{NV}$ is measured to be \SI{\sim 120}{kHz}.
Residual transverse magnetic fields are observed to shift the nuclear spin transition frequencies, especially for $^{15}\mathrm{NV}$.
We have precisely determined the set of parameters relevant for the development of nuclear-spin-based diamond quantum sensors with greatly reduced sensitivity to environmental factors.

\end{abstract}

\maketitle

\section{Introduction}

In recent years, color centers in diamond, and in particular, the nitrogen-vacancy (NV) center, have emerged as one of the key  platforms for quantum-technology applications, particularly in sensing \cite{DEG2017}. As the technology matures, detailed knowledge of the parameters of the system and their environmental dependence become a prerequisite of development of accurate devices such as magnetometers, gyroscopes, clocks, as well as multisensors. For NV-based rotation sensing \cite{LED2012, AJO2012, MAC2012}, one may use $^{14}\mathrm{NV}$ and $^{15}\mathrm{NV}$ centers, where using two isotopes is important for differential measurements to separate rotational, magnetic, and temperature effects \cite{KOR2005, DON2013, Walker2016, NMRgyro2018, NMRgyro2019, SORENSEN2020}.

In this work, which follows the earlier experimental studies of the temperature dependence of the ground state zero-field splitting parameter $D$ \cite{ACO2010}, the electric quadrupole hyperfine splitting parameter $Q$ \cite{Jarmola2020Robust} and the theoretical analysis \cite{DOH2014PRB}, we present a complete experimental characterization of temperature dependence of the coupling parameters of NV centers for both nitrogen isotopes ($^{14}\mathrm{NV}$ and $^{15}\mathrm{NV}$), including the dependence of the magnetic hyperfine coupling parameters $A_{||}$ and $A_{\perp}$. We find that NV parameters are generally temperature dependent, with relative sensitivity ranging from 7 to 90\,ppm/K at room temperature, depending on the parameter.

Additional findings of this work include identification of the relatively high sensitivity of $^{15}\mathrm{NV}$ nuclear spin levels to misalignment of the magnetic field to the NV axis, as well as measurement of the isotopic shift in $D$ at a level of \SI{\sim 40}{ppm}. The latter is important for testing the theoretical models of the system in order to attain a level of understanding necessary for accurate modeling of devices. Note that $^{15}\mathrm{NV}$ nuclear spins were recently explored as a resource for quantum sensing not relying on microwave or radio-frequency fields in \cite{Burgler2022}.

\begin{figure*}
\centering
    \includegraphics[width=1.0\columnwidth]{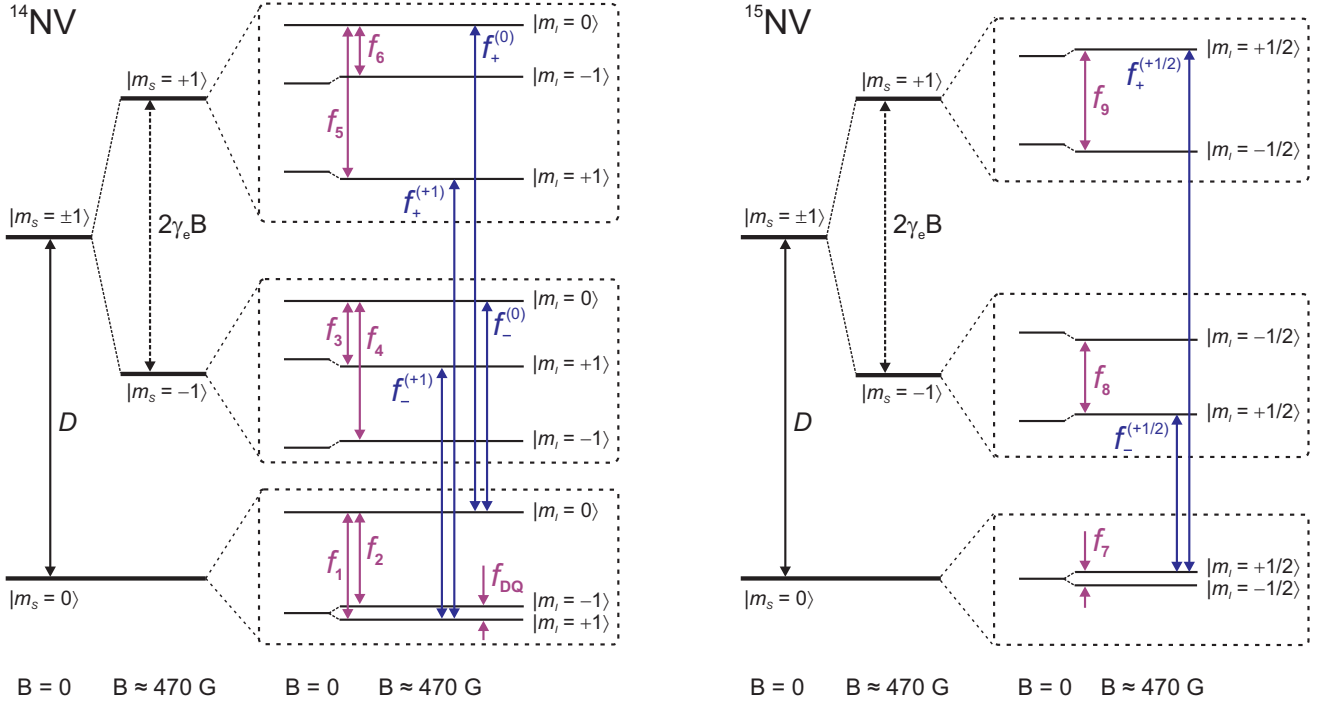}
    \caption{\label{fig:EnergyLevels} \textbf{Energy level diagrams for the electronic ground states of $^{14}\mathrm{NV}$ and $^{15}\mathrm{NV}$.} Energy levels are described by electronic spin ($m_s$) and nuclear spin ($m_I$) quantum numbers. The electron-spin transitions used in this experiment are shown with blue arrows, and are labeled as $f_\pm^{(m_I)}$ for $\ket{m_s=0}\leftrightarrow \ket{m_s=\pm1}$, where $m_I$ denotes the nuclear-spin state of the transition. The nuclear-spin transitions are shown with purple arrows and are labeled $f_1$ to $f_9$.
    }
\end{figure*}

\begin{figure*}[t]
\centering
    \includegraphics[width=1.0\columnwidth]{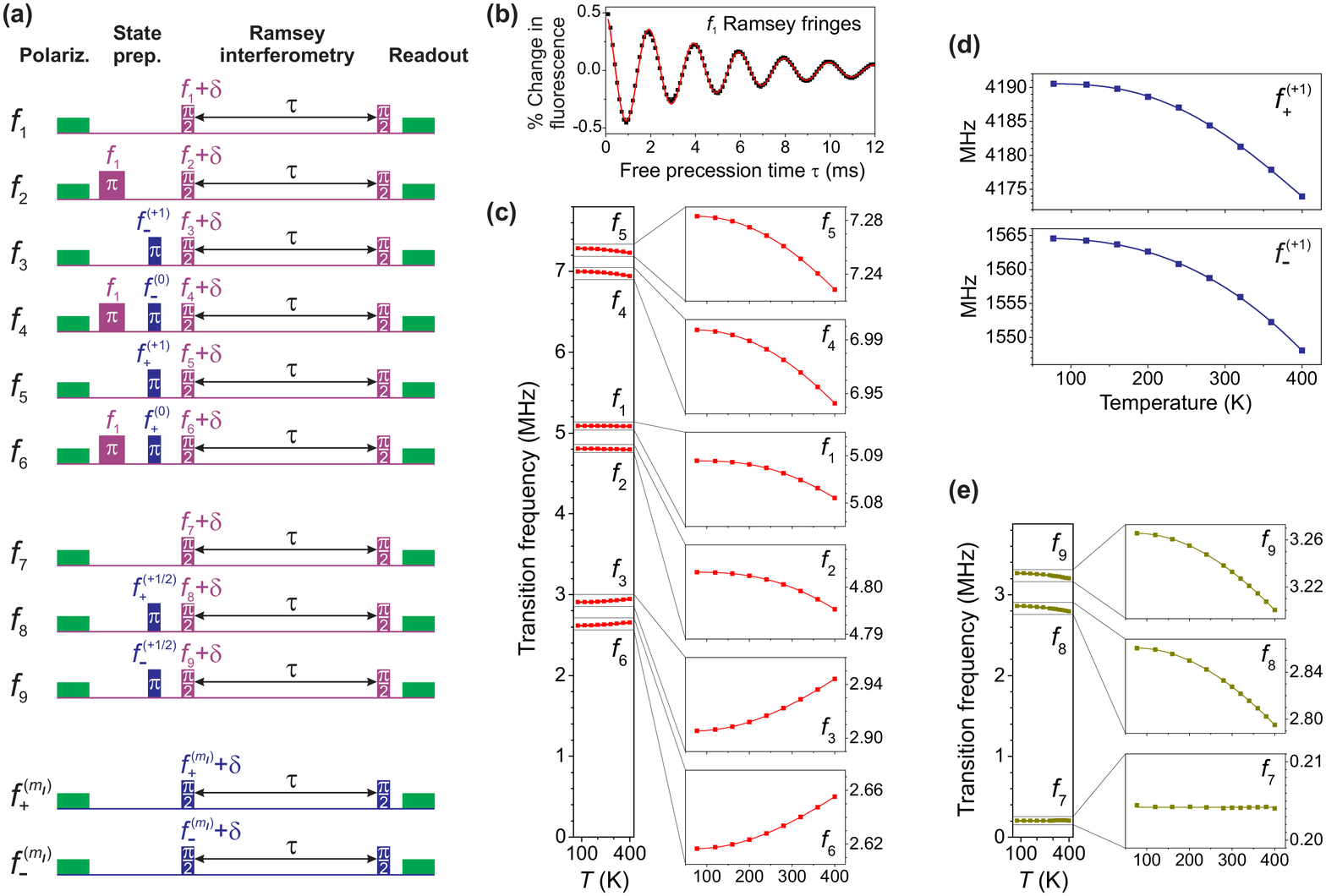}
    \caption{\label{fig:RamseyMeasurements}
    \textbf{Ramsey measurements of ground state transition frequencies.}
    \textbf{(a)} Nuclear-spin transition frequencies $f_1$ to $f_6$ ($^{14}\mathrm{NV}$) and $f_7$ to $f_9$ ($^{15}\mathrm{NV}$) as well as electron-spin transition frequencies $f_+$ and $f_-$ for both isotopes are measured using Ramsey interferometry. After optical polarization (green) with green light, the nuclear spin is manipulated with a series of RF (purple) and MW (blue) pulses to prepare a superposition of the two relevant energy states. The superposition then precesses at the detuning frequency $\delta$ for a variable time $\tau$, after which a $\pi/2$ RF pulse converts the acquired phase into a population difference to be read out optically.
    \textbf{(b)} Example of the nuclear Ramsey interferometry measurement. The oscillation frequency of the Ramsey fringes corresponds to the detuning $\delta$ from the transition frequency $f_1$.
    \textbf{(c)} Temperature dependence of $f_1$ to $f_6$ for sample G2 at $B \approx \SI{470}{G}$. The y-axis range for $f_1$ and $f_2$ subplots has been reduced ($\SI{70}{kHz} \rightarrow \SI{20}{kHz}$) to show the reduced temperature dependence of $f_1$ and $f_2$.
    \textbf{(d)} Temperature dependence of $f_+$ and $f_-$ for $^{14}\mathrm{NV}$ ($^{15}\mathrm{NV}$ not shown) for sample G2 at $B \approx \SI{470}{G}$.
    \textbf{(e)} Temperature dependence of $f_7$ to $f_9$ for sample M1 at $B \approx \SI{468}{G}$. The y-axis range for the $f_7$ subplot has been reduced ($\SI{80}{kHz} \rightarrow \SI{12}{kHz}$).
   }
\end{figure*}

\section{\label{sec:Background}Theoretical Background}

The Hamiltonian for the electronic ground state of $^{14}\mathrm{NV}$ and $^{15}\mathrm{NV}$ \cite{DOH2013} is given by:
\begin{equation}
\begin{split}
    H &= D S_z^2 + Q I_z^2 + A_{||} S_z I_z + \gamma_e B \, S_z - \gamma_n B \, I_z \\
 &+ \frac{A_{\perp}}{2} \left(S_+ I_- + S_- I_+\right) \label{eq:Hamiltonian},
\end{split}
\end{equation}
where
$D$ is the ground state zero-field splitting parameter of the NV center, $Q$ is the nuclear electric quadrupole parameter (only for $^{14}\mathrm{NV}$), $A_{||}$ and $A_{\perp}$ are the longitudinal and transverse magnetic hyperfine coupling parameters (see Table~\ref{table:297Parameters} for parameter values at \SI{297}{K}), $\gamma_e$ is the gyromagnetic ratio of the NV center ($\SI{2.8033 +- 0.0003}{MHz \per G}$ \cite{FEL2009}), $\gamma_n$ is the gyromagnetic ratio of the nitrogen nuclear spin ($^{14}\gamma_n = \SI{307.59+-0.03}{Hz/G}$, $^{15}\gamma_n = \SI{-431.50+-0.04}{Hz/G}$), $\textbf{B}$ is the external magnetic field applied along the z-axis (NV symmetry axis), and $\textbf{S}$ and $\textbf{I}$ are electron and nuclear spin operators, respectively.

The energy level diagrams for the electronic ground states of $^{14}\mathrm{NV}$ and $^{15}\mathrm{NV}$ are shown in Fig.~\ref{fig:EnergyLevels}.
The electron-spin transitions $\ket{m_s=0} \leftrightarrow \ket{m_s=\pm 1}$ are labeled as $f_\pm^{(m_I)}$, where $m_I$ denotes the nuclear-spin state.
The $^{14}\mathrm{NV}$ and $^{15}\mathrm{NV}$ nuclear-spin transitions are labeled $f_1$ to $f_6$ and $f_7$ to $f_9$, respectively, according to the diagram.
The nuclear-spin double quantum transition with frequency $f_1-f_2$ is labeled as $f_{DQ}$ and is of particular interest for rotation sensing \cite{Jarmola2021,Soshenko2021} and comagnetometry \cite{Chu2022, BAR2020}.

Nuclear-spin transition frequencies for both $^{14}\mathrm{NV}$ and $^{15}\mathrm{NV}$ can be derived using perturbation theory (see Appendix~\ref{sec:Appendix:PerturbationTheory}), and are described to lowest order in $A_{\perp}/(D\pm\gamma_e B)$ by the following expressions:
\begin{alignat}{4}
\label{eq:EquationsN14}
    f_1 &\approx |Q|            &{}+ {}^{14}\gamma_n B &- \frac{A_{\perp}^2}{D-\gamma_e B} &  \nonumber \\
    f_2 &\approx |Q|            &{}- {}^{14}\gamma_n B &- \frac{A_{\perp}^2}{D+\gamma_e B} &  \nonumber \\
    f_3 &\approx |Q| - |A_{||}| &{}+ {}^{14}\gamma_n B &                                   &  \nonumber \\
    f_4 &\approx |Q| + |A_{||}| &{}- {}^{14}\gamma_n B &+ \frac{A_{\perp}^2}{D-\gamma_e B} &  \nonumber \\
    f_5 &\approx |Q| + |A_{||}| &{}+ {}^{14}\gamma_n B &+ \frac{A_{\perp}^2}{D+\gamma_e B} &  \nonumber \\
    f_6 &\approx |Q| - |A_{||}| &{}- {}^{14}\gamma_n B &                                   &
\end{alignat}

\begin{alignat}{4}
\label{eq:EquationsN15}
    f_7 &\approx  &{}\phantom{-}{}&|^{15}\gamma_n| B &{}+{}&\frac{A_{\perp}^2}{2}\left(\frac{1}{D - \gamma_e B} -\frac{1}{D + \gamma_e B}\right) \nonumber \\
    f_8 &\approx A_{||} &{}-{}&|^{15}\gamma_n| B &{}-{}&\frac{A_{\perp}^2}{2}\left(\frac{1}{D - \gamma_e B} \right) \nonumber \\
    f_9 &\approx A_{||} &{}+{}& |^{15}\gamma_n| B &{}-{}&\frac{A_{\perp}^2}{2}\left(\frac{1}{D + \gamma_e B}\right) \, .
\end{alignat}

In this work, the transition frequencies $f_1$ to $f_9$, $f_\pm^{(+1)}$, and $f_\pm^{(+1/2)}$ are measured in the presence of an axial field ($B \approx 470\,\mathrm{G}$) at temperatures ranging from \SI{77}{K} to \SI{400}{K}.

\renewcommand{\arraystretch}{1.2}
\begin{table*}
    \caption{\textbf{Diamond samples.}
        The estimated concentrations of substitutional nitrogen $\left[\mathrm{N}\right]$, NV centers $\left[\mathrm{NV}\right]$, and ${}^{13}\mathrm{C}$ atoms $\left[^{13}\mathrm{C}\right]$, in addition to the nitrogen isotopic ratio $\left[^{14}\mathrm{N}\right] : \left[{}^{15}\mathrm{N}\right]$,  electron-spin dephasing time $T_2^*$, and electron-spin coherence time $T_2$ are listed for each sample used in experiments. Diamonds were grown using chemical vapor deposition and were obtained from Element Six. Electron-spin $T_2^*$ was measured using Ramsey interferometry, and $T_2$ was measured using Hahn echo techniques. 
    }
    \label{table:Samples}
    \begin{ruledtabular}
    \begin{tabular}{cddccdc}
    Sample Name &  
    \multicolumn{1}{c}{\textrm{$\left[\mathrm{N}\right]$ (ppm)}}     &
    \multicolumn{1}{c}{\textrm{$\left[\mathrm{NV}\right]$ (ppm)}}     &
    \multicolumn{1}{c}{\textrm{$\left[^{14}\mathrm{N}\right] : \left[{}^{15}\mathrm{N}\right]$}} &
    \multicolumn{1}{c}{\textrm{$\left[{}^{13}\mathrm{C}\right]$ (\%)}}     &
    \multicolumn{1}{c}{\textrm{$T_2^*$ (\si{\micro s})}}     &
    \multicolumn{1}{c}{\textrm{$T_2$ (\si{\micro s})}}     \\[0.5 ex]
    \colrule
    F7 & \sim 0.1 & \sim 0.01 & 99.6 : 0.4 & 1.1 & 1.8(2) & 600(20) \\
    G2 &  10   & 4  & 99.6 : 0.4 & $<$ 0.01 &  1.0(2) &  11(1) \\
    M1 &  \sim 10   & \sim 1  & 20 : 80    & - & 0.25(2) & 15(2) \\
    G3 &  16   & 4.5  & 50 : 50    & 1.1 &  0.35(3) &  22(2)
    \end{tabular}
    \end{ruledtabular}
\end{table*}

\section{\label{sec:Methods}Experimental Methods}

We used a custom-built epifluorescence microscopy setup to measure optically detected magnetic resonances (ODMRs) in an ensemble of NV centers. 
Four diamond samples, whose properties are listed in Table \ref{table:Samples}, were used in our experiments: two with natural isotopic ratio of nitrogen and two with enhanced $^{15}\mathrm{N}$ concentration.

The diamond sample was mounted inside a continuous-flow microscopy cryostat (Janis ST-500). A bias magnetic field $B$ (\SIrange{470}{480}{G}) was applied along one of the NV axes using two temperature-compensated samarium-cobalt (SmCo) ring magnets, which were arranged in a Helmholtz-like configuration that minimizes magnetic field gradients across the detected volume. A 0.79–numerical aperture aspheric condenser lens was used to illuminate a spot on the diamond with $\SI{\sim 30}{mW}$ of 532 nm laser light and collect fluorescence. The fluorescence was separated from the excitation light by a dichroic mirror, passed through a band-pass filter (650 to 800 nm), and detected by a free-space Si photodiode. Microwave (MW) and radio-frequency (RF) signals were delivered using a 160\,$\upmu$m diameter copper wire placed on the diamond surface next to the optical focus. 

The transition frequencies of the NV ground state were measured with Ramsey interferometry using pulse sequences that are shown in Fig.~\ref{fig:RamseyMeasurements}(a). 
For each nuclear-spin transition frequency ($f$) between a pair of states ($\ket{m_s,m_I}$ and $\ket{m_s,m_I'}$), the following steps are performed: optical polarization, state preparation, Ramsey interferometery, and optical readout. Polarization is done with a green laser pulse with duration \SIrange{100}{200}{\micro s}, which polarizes the electron and nuclear spins into $\ket{0, +1}$ (or $\ket{0,+1/2}$ for $^{15}\mathrm{NV}$)
\cite{JAC2009, SME2009, STE2010PRB, Jarmola2020Robust}.
State $\ket{m_s,m_I}$ is prepared by transferring population
using a sequence of RF and MW $\pi$ pulses.
Next, a superposition of the states $\ket{m_s,m_I}$ and $\ket{m_s,m_I'}$ is created using a Ramsey $\pi/2$ pulse with frequency $f_{RF}$ (duration \SIrange{10}{100}{\micro s}).
The superposition then accumulates a phase at a rate given by the detuning frequency, $\delta = f_{RF}-f$ (typically \SIrange{0.5}{10}{kHz}).
After a variable delay time $\tau$, the acquired relative phase is projected into a population difference with a second $\pi/2$ pulse and read out optically \cite{Jarmola2020Robust}.
To determine $f$, we obtain $\delta$ by fitting the Ramsey oscillations with an exponentially decaying sinusoidal function.
Measurements of electron-spin transition frequencies ($f_+$, $f_-$) were performed in the same way, using MW pulses instead of RF pulses.
Figure~\ref{fig:RamseyMeasurements}(b) shows an example of a Ramsey measurement for the nuclear transition frequency $f_1$.

\section{\label{sec:Results}Results}

The transition frequencies of $^{14}\mathrm{NV}$ ($f_1$ to $f_6$, $f_+^{(+1)}$,~$f_-^{(+1)}$) and $^{15}\mathrm{NV}$ ($f_7$ to $f_9$, $f_+^{(+1/2)}$,~$f_-^{(+1/2)}$) were measured using the previously described Ramsey interferometry technique [Fig.~\ref{fig:RamseyMeasurements}(a,b)] in the presence of an axial magnetic field $B \approx \SI{470}{G}$ for temperatures ranging from  $\SI{77}{K}$ to $\SI{400}{K}$ using four samples; see Table~\ref{table:Samples}.

Figure~\ref{fig:RamseyMeasurements}(c,d) shows the temperature dependence of the nuclear-spin and electron-spin transitions in $^{14}\mathrm{NV}$ for sample G2.
Transition frequencies $f_1$ and $f_2$ are least sensitive to temperature, varying by $\SI{8}{kHz}$ across the range, while $f_3$, $f_6$ and $f_4$, $f_5$ vary by $\SI{\sim 40}{kHz}$ and $\SI{\sim 50}{kHz}$, respectively.
These temperature dependences are largely determined by the temperature dependence of $Q$ and $A_{||}$, according to Eq.~(\ref{eq:EquationsN14}).
Figure~\ref{fig:RamseyMeasurements}(e) shows the temperature dependence of the nuclear-spin transitions in $^{15}\mathrm{NV}$ for sample M1, which has been isotopically enriched with $^{15}\mathrm{N}$.
Transition frequencies $f_8$ and $f_9$ were measured to vary by $\SI{\sim 70}{kHz}$ across the measured range of temperatures, while the temperature dependence of $f_7$ was found to be three orders of magnitude smaller; see Sec~\ref{sec:Discussion} for detailed analysis.

\begin{figure*}
\centering
    \includegraphics[width=1.0\columnwidth]{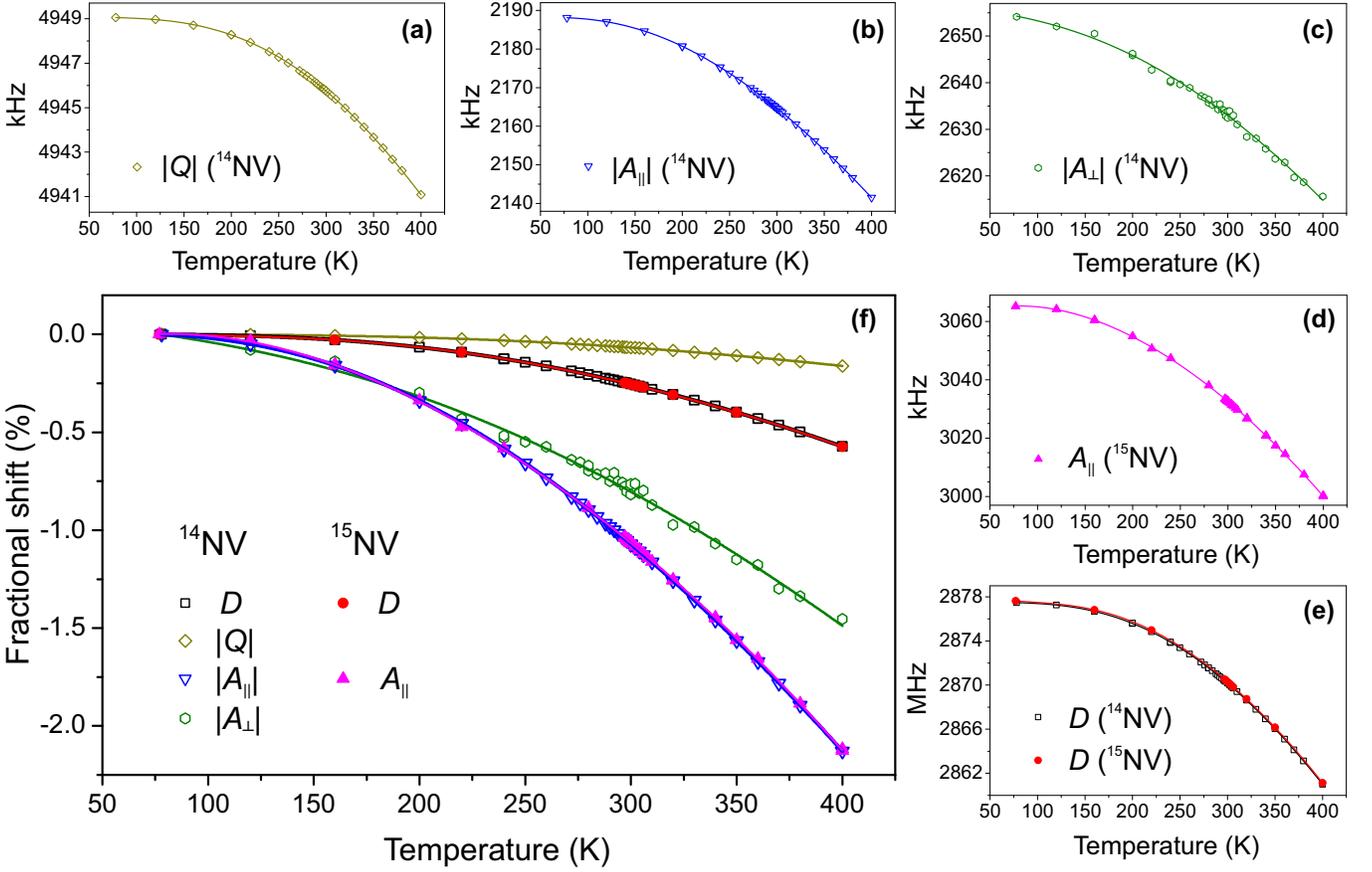}
    \caption{\label{fig:CouplingTemperatureData} \textbf{Temperature dependence of coupling parameters. (a)--(e)}
    Coupling parameters $D$, $Q$, $A_{||}$, and $A_{\perp}$ were extracted numerically (see Appendix~\ref{sec:Appendix:Numerical}) from measured nuclear-spin and electron-spin transition frequencies.
    Coupling parameters are plotted against temperature, for both $^{14}\mathrm{NV}$ and $^{15}\mathrm{NV}$, using data from all diamond samples.
    The solid lines are fourth-degree polynomial fits.
    \textbf{(f)} Fractional temperature dependence of all parameters whose data is presented in \textbf{(a)--(e)}. $^{14}\mathrm{NV}$ and $^{15}\mathrm{NV}$ were found to have similar fractional temperature shifts in $D$ and in $A_{||}$.
   }
\end{figure*}

Using all measured transition frequencies, we apply numerical methods described in Appendix~\ref{sec:Appendix:Numerical} to extract values for the coupling parameters for both $^{14}\mathrm{NV}$ ($D$, $Q$, $A_{||}$, $A_{\perp}$, $\gamma_e / ^{14}\gamma_n$) and $^{15}\mathrm{NV}$ ($D$, $A_{||}$, $A_{\perp}$, $\gamma_e / ^{15}\gamma_n$). 

\renewcommand{\arraystretch}{1.2}
\begin{table*}[!t]
\caption{\textbf{Experimentally determined coupling parameters at 297\,K.}
    The values and temperature derivatives of coupling parameters $^{14}\mathrm{NV}$ ($D$, $Q$, $A_{||}$, $A_{\perp}$) and $^{15}\mathrm{NV}$ ($D$, $A_{||}$) at \SI{297}{K} are obtained from the polynomial fits of the temperature dependences shown in Fig.~\ref{fig:CouplingTemperatureData}. $^{15}\mathrm{NV}$ $A_{\perp}$ is obtained by scanning the transverse magnetic field with coils; see Fig.~\ref{fig:Fractional_DQf7}(c).
}
\label{table:297Parameters}
\begin{ruledtabular}
\begin{tabular}{ccz{5,10}z{4,7}z{3,4}z{2,11}}
Isotope & Parameter                                & 
\multicolumn{1}{c}{\textrm{Value}}                 &
\multicolumn{1}{c}{\textrm{\nth{1} Derivative}}    &
\multicolumn{1}{c}{\textrm{Fractional Derivative}} &
\multicolumn{1}{c}{\textrm{\nth{2}  Derivative}}   \\
&             &
\multicolumn{1}{c}{\textrm{\si{kHz}}}          &
\multicolumn{1}{c}{\textrm{\si{Hz \per K}}}    &
\multicolumn{1}{c}{\textrm{\si{ppm \per K}}}   &
\multicolumn{1}{c}{\textrm{\si{Hz \per K^2}}}  \\
\colrule
$^{14}\mathrm{NV}$
& $D_{\phantom{0}}$ &  2870.28\rlap{$(3) \times 10^3$}
                    & -72.5   \rlap{$(5) \times 10^3$}
                    & -25.3   \rlap{$(2)$}     
                    & -0.39   \rlap{$(1) \times 10^3$}\\
& $Q_{\phantom{0}}$ & -4945.88\rlap{$(1)$} &  35.5\rlap{$(3)$} &  -7.17\rlap{$(6)$} &  0.22 \rlap{$(1)$} \\
& $A_{||}         $ & -2165.19\rlap{$(8)$} & 197  \rlap{$(1)$} & -91.0 \rlap{$(5)$} &  0.73 \rlap{$(6)$} \\
& $A_{\perp}      $ & -2635   \rlap{$(2)$} & 154  \rlap{$(5)$} & -58   \rlap{$(2)$} &  0.53 \rlap{$(3)$} \\
\colrule
$^{15}\mathrm{NV}$
& $D_{\phantom{0}}$ & 2870.38\rlap{$(3) \times 10^3$}
                    &  -72   \rlap{$(1) \times 10^3$}
                    &  -25.1 \rlap{$(3)$}
                    &   -0.40\rlap{$(2) \times 10^3$}\\
& $A_{||}         $ & 3033.3\rlap{$(1)$}   & -269   \rlap{$(3)$} & -89   \rlap{$(1)$} & -0.98 \rlap{$(8)$} \\
& $A_{\perp}      $ & 3680  \rlap{$(20)$}  &                     &                    & \\
\end{tabular}
\end{ruledtabular}
\end{table*}

\begin{figure*}
\centering
    \includegraphics[width=1.0\columnwidth]{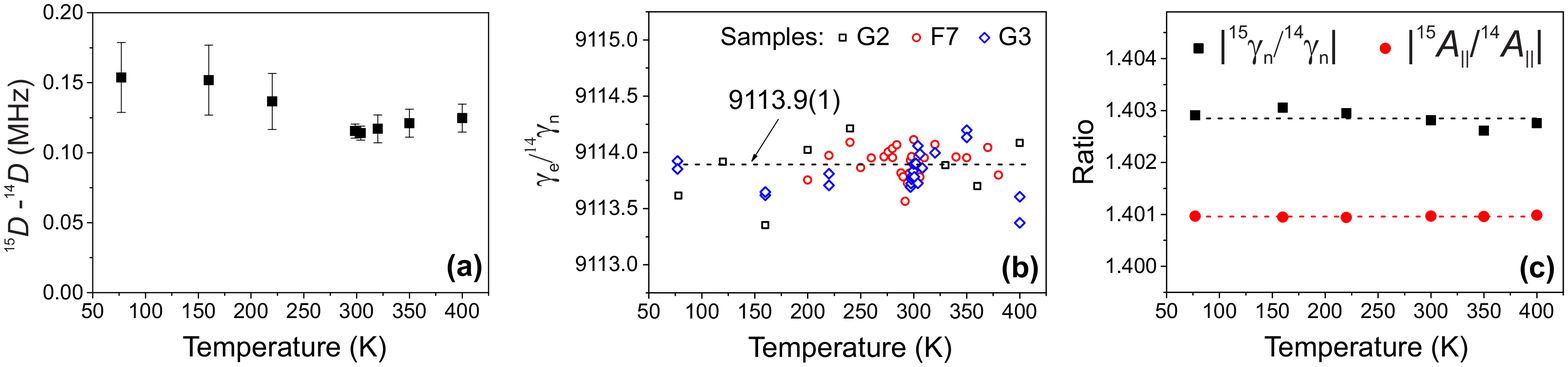}
    \caption{\label{fig:TemperatureIndependentResults} \textbf{Temperature-insensitive parameters.}
    \textbf{(a)}~Difference in the zero-field splitting parameter $D$ between $^{14}\mathrm{NV}$ and $^{15}\mathrm{NV}$. \textbf{(b)}~Ratio of electron-spin gyromagnetic ratio $\gamma_e$ and $^{14}\mathrm{NV}$ nuclear-spin gyromagnetic ratio $^{14}\gamma_n$.
    \textbf{(c)}~Isotopic ratios of $\gamma_n$ and of $A_{||}$ for  $^{14}\mathrm{NV}$ and $^{15}\mathrm{NV}$.
    Markers -- experimental data, dashed lines -- mean values.
    In \textbf{(a)} and \textbf{(c)} the data were measured using sample G3, which has a $50:50$ ratio of $[^{15}\mathrm{NV}]:[^{14}\mathrm{NV}]$.
    }
\end{figure*}

Numerically extracted coupling parameters for $^{14}\mathrm{NV}$ ($D$, $Q$, $A_{||}$, $A_{\perp}$) and $^{15}\mathrm{NV}$ ($D$, $A_{||}$) are plotted as a function of temperature in Fig.~\ref{fig:CouplingTemperatureData}(a-e).
Each parameter's temperature dependence was fitted to a 4th order polynomial.
Figure~\ref{fig:CouplingTemperatureData}(f) shows the fractional shift of each parameter as a function of temperature.
We observe $^{14}\mathrm{NV}$ and $^{15}\mathrm{NV}$ to have identical fractional dependence for both $D$ and $A_{||}$.

In the case of $^{15}\mathrm{NV} A_{\perp}$, the following factors prevented the determination of its temperature dependence:
(i) $A_{\perp}$ has a weak contribution to the nuclear-spin transition frequencies,
(ii) the $^{15}\mathrm{NV}$ nuclear-spin transition frequencies ($f_7$ in particular) are sensitive to magnetic field misalignment $\theta$; see Sec~\ref{sec:TemperatureAngle},
(iii) $^{15}\mathrm{NV}$ only has three nuclear-spin transitions, whose frequencies are determined by four parameters: $A_{||}$, $A_{\perp}$, $^{15}\gamma_n B$, $\theta$.
We were able to overcome these issues at room temperature by scanning the transverse magnetic field with coils [see Fig.~\ref{fig:Fractional_DQf7}(c)] and obtained $A_{\perp} = \SI{3.68+-0.02}{MHz}$, which is in agreement with \cite{FEL2009}.
In future studies, this approach can be used to obtain the temperature dependence of $^{15}\mathrm{NV} A_{\perp}$.

Table~\ref{table:297Parameters} lists the values and temperature derivatives of each coupling parameter at \SI{297}{K} for both $^{14}\mathrm{NV}$ and $^{15}\mathrm{NV}$, obtained from the polynomial fits.
The room temperature value for each measured parameter is consistent with previously reported values \cite{FEL2009,SME2009,STE2010PRB,CHEN2015}.
Temperature derivatives are also consistent with previously reported values for $^{14}\mathrm{NV}$ parameters $D$ \cite{ACO2010,Che2011,DU2022}, $Q$ \cite{Jarmola2020Robust,Soshenko2020,Wang2022,DU2022}, and $A_{||}$ \cite{Soshenko2020,Wang2022,DU2022}. Recent theoretical work \cite{Tang2022} provided \textit{ab initio} evaluation of the temperature dependences of several parameters for $^{14}$NV system. In the cases where the same parameters were calculated in \cite{Tang2022} and measured here, we find good agreement.
To the best of our knowledge, temperature derivatives for $^{14}\mathrm{NV}$ $A_{\perp}$ and $^{15}\mathrm{NV}$ $A_{||}$ are reported here for the first time.

We observe an isotopic shift of \SI{\sim 120}{kHz} (\SI{\sim 40}{ppm}) in the $D$ parameter, with $D$ larger for $^{15}\mathrm{NV}$. Figure~\ref{fig:TemperatureIndependentResults}(a) shows the shift in $D$ across the full range of measured temperatures for the G3 sample, which has a $50\!:\!50$ isotopic ratio. This effect is also clearly visible as a difference in separation of peaks in the ODMR signal (see Appendix~\ref{sec:Appendix:ODMR}). The relatively small isotopic effect in $D$ is in line with the fact that even a replacement of the species adjacent to the vacancy changes the zero-field splitting only slightly, see recent work \cite{Umeda2022}, where the $D$ values were reported to be 2888\,MHz and 2913\,MHz for the OV$^0$ and BV$^-$, respectively.

Figure~\ref{fig:TemperatureIndependentResults}(b) shows the numerically extracted values of $\gamma_e / ^{14}\gamma_n$
for three different samples. The mean value of $\gamma_e / ^{14}\gamma_n$ is measured to be $\num{9113.9 +- 0.1}$, which is in agreement with \cite{Du2021PRL}.
Using the literature value for $\gamma_e = \SI{2.8033 +- 0.0003}{MHz \per G}$ \cite{FEL2009}, this corresponds to $^{14}\gamma_n = \SI{307.59 +- 0.03}{Hz \per G}$.
The value for $\gamma_e / ^{14}\gamma_n$ can also be approximated without numerical methods from the measured frequencies directly [see Eq.~(\ref{eq:EquationsN14})]:
\begin{equation}
    \frac{\gamma_e}{^{14}\gamma_n} \approx \frac{f_+^{(0)} - f_-^{(0)}}{f_3 - f_6} \approx \frac{f_+^{(+1)} - f_-^{(+1)} + f_5 - f_3}{f_3 - f_6}\,.
\end{equation}
Figure~\ref{fig:TemperatureIndependentResults}(c) shows the isotopic ratio of $ \gamma_n$
$\big(\left|^{15}\gamma_n/^{14}\gamma_n\right| = \num{1.40285 +- 0.00006}$, in agreement with \cite{NMR2005}\big) and the isotopic ratio of $A_{||}$
$\left(\left|^{15}A_{||}/^{14}A_{||}\right| = \num{1.40096 +- 0.00001}\right)$
obtained from measurements using the G3 sample. 
The difference in these ratios can be used to verify theoretical models.
When using the literature value for $\gamma_e$, we obtain $^{15}\gamma_n = \SI{-431.50 +- 0.04}{Hz \per G}$.

\section{Discussion}
\label{sec:Discussion}
\subsection{Temperature and angular dependence of $f_{DQ}$ and $f_7$}
\label{sec:TemperatureAngle}

\begin{figure}[!h]
\centering
    \includegraphics[width=1\columnwidth]{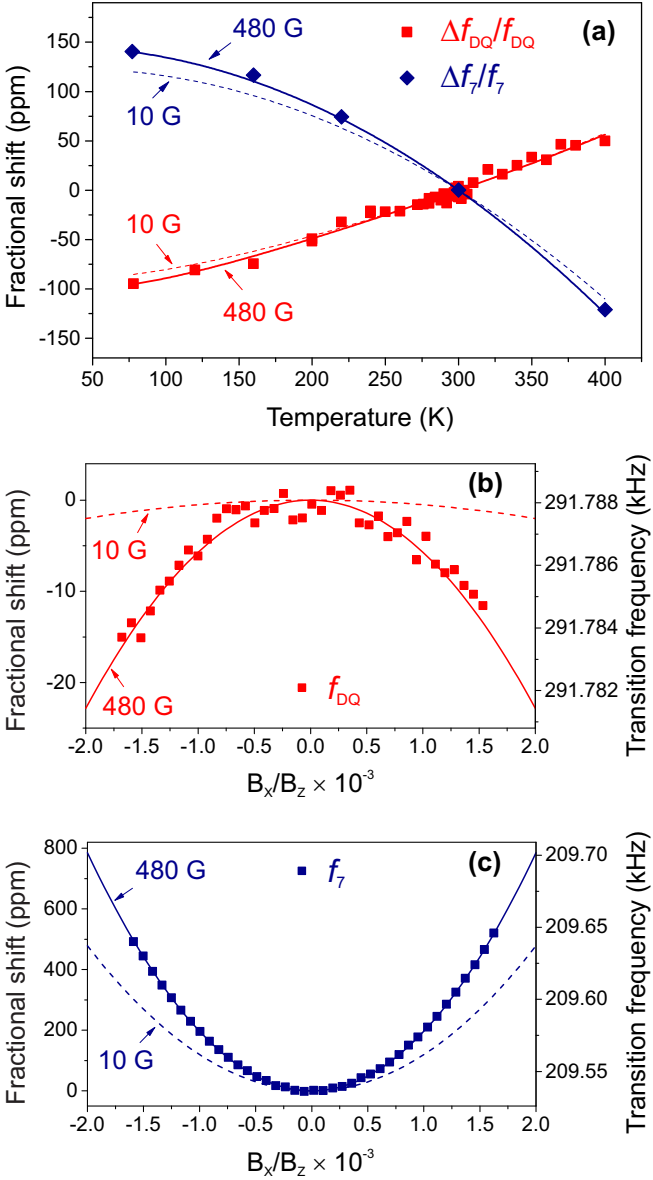}
    \caption{\label{fig:Fractional_DQf7} \textbf{Temperature and angular dependence of $f_{DQ}$ and $f_{7}$.}
    \textbf{(a)} The fractional shifts in nuclear transition frequencies $f_{DQ}$ (red) and $f_{7}$ (blue) are plotted as a function of temperature.
    Markers represent experimental data after correcting for variations in the magnetic field between measurements.
    Solid and dashed lines were obtained from Eqs.~(\ref{eq:EquationsDQ}),(\ref{eq:Equationsf7}) at \SI{480}{G} and \SI{10}{G}, respectively. 
    \textbf{(b),(c)} The fractional shifts in nuclear transition frequencies $f_{DQ}$ (red) and $f_{7}$ (blue), respectively, are plotted as a function of magnetic field misalignment with respect to the NV axis for fixed values of $B_z$.
    Markers represent experimental data, and solid (\SI{480}{G}) and dashed (\SI{10}{G}) lines were obtained by numerically diagonalizing the Hamiltonian [Eq.~(\ref{eq:Hamiltonian})] using values from Table~\ref{table:297Parameters}.
    For $f_7$ (c), $A_{\perp}$ is treated as a free parameter, and fit to the experimental data in order to obtain $A_{\perp} = \SI{3.68 +- 0.02}{MHz}$.
       }
\end{figure}

The nuclear-spin transitions $f_{DQ}$ and $f_{7}$ within the $m_s = 0$ manifold of $^{14}\mathrm{NV}$ and $^{15}\mathrm{NV}$ are of particular interest for sensing applications such as rotation sensing \cite{Jarmola2021, Soshenko2021} and comagnetometry \cite{Chu2022, BAR2020}.
Precise knowledge of their transition frequencies and their dependence on environmental factors (temperature, magnetic field) is essential for optimal sensor performance.
The transition frequencies $f_{DQ}$ and $f_{7}$ can be obtained from Eqs.~(\ref{eq:EquationsN14}),(\ref{eq:EquationsN15}) and are described by the following expressions:
\begin{alignat}{4} \label{eq:EquationsDQ}
    f_{DQ} &\approx  &{}\phantom{-}{}&2\cdot{}^{14}\gamma_n & B & \left(1 - \left|\frac{\gamma_e}{^{14}\gamma_n}\right|\frac{A_{\perp}^2}{D^2 - \gamma_e^2 B^2}\right)\\ \label{eq:Equationsf7}
    f_7 &\approx  &{}\phantom{-}{}&|^{15}\gamma_n| & B & \left(1 + \left|\frac{\gamma_e}{^{15}\gamma_n}\right|\frac{A_{\perp}^2}{D^2 - \gamma_e^2 B^2}\right).
\end{alignat}
The frequencies of these transitions do not depend on the coupling parameters $Q$, $A_{||}$ and are determined primarily by the nuclear Zeeman shift $\Delta m_I\gamma_n B$, resulting in a greatly reduced temperature dependence compared to other nuclear spin transitions.
Therefore, measurements of the temperature dependences of $f_{DQ}$ and $f_{7}$ require more precise control of the bias magnetic field.
Over the range of temperatures used in this experiment, the bias magnetic field varied by $\sim \SI{1}{G}$, primarily due to thermal expansion of the sample holder in the presence of magnetic field gradients.
This variation in the bias magnetic field was measured using electron-spin transition frequencies and subsequently used to obtain corrected values for $f_{DQ}$ and $f_{7}$ corresponding to \SI{480}{G}.
When the temperature is changed from 77K to 400K, the transition frequencies $f_{DQ}$ and $f_{7}$ are observed to shift by \SI{140}{ppm} (\SI{44}{Hz} at \SI{480}{G}) and \SI{-260}{ppm} (\SI{-55}{Hz} at \SI{480}{G}), respectively; see Fig.~\ref{fig:Fractional_DQf7}(a). 
This corresponds to fractional temperature derivatives of \SI{0.52(1)}{ppm/K} (\SI{0.15}{Hz/K}) for $f_{DQ}$ and \SI{-1.1(1)}{ppm/K} (\SI{-0.10}{Hz/K}) for $f_{7}$ at 297K.

The temperature dependence of $f_{DQ}$ and $f_{7}$ arises from the temperature dependence of $A_{\perp}^2/D^2$ and is described by Eqs.~(\ref{eq:EquationsDQ}),(\ref{eq:Equationsf7}).
These equations are used with experimentally obtained polynomial fits of $A_{\perp}$ and $D$ (see Fig.~\ref{fig:CouplingTemperatureData}), to generate the solid and dashed lines in Fig.~\ref{fig:Fractional_DQf7}(a).
For $f_{7}$, $A_{\perp}$ is assumed to have the same fractional dependence on temperature as $A_{||}$.
For sufficiently small fields $\gamma_e B << D$ (i.e., $B = \SI{10}{G}$), the transition frequencies are approximately linear in the magnetic field, and thus the fractional shift is independent of magnetic field.

Magnetic field misalignment (from the NV axis) is another factor that can significantly shift the transition frequencies of $f_{DQ}$ and $f_7$, and must be considered in sensing applications.
Angular-dependent shifts in $f_7$ are significantly stronger than shifts in $f_{DQ}$ because of the lack of a stabilizing coupling parameter (i.e., $Q$) in the effective nuclear-spin Hamiltonian.
The frequency shifts $f_{DQ}$ and $f_{7}$ exhibit a quadratic dependence on misalignment angle, and are shown in Figs.~\ref{fig:Fractional_DQf7}(b) and \ref{fig:Fractional_DQf7}(c), respectively.
The alignment of the magnetic field was controlled using two pairs of coils oriented perpendicular to the bias magnetic field $B_z$.
The transverse magnetic field $B_x$ was precisely determined using the NV electron spin transitions of the three non-axial NV sub-ensembles.
We measure a frequency shift of \SI{5.0}{Hz} in $f_{DQ}$ and of \SI{130}{Hz} in $f_7$ when misaligning the magnetic field by $\theta = \SI{0.1}{\degree}$ ($B_x \approx \SI{0.8}{G}$) at a field of $B_z = \SI{480}{G}$.

We use numerical methods (Appendix~\ref{sec:Appendix:Numerical}) together with experimental values from Table~\ref{table:297Parameters} to obtain theoretical predictions for $f_{DQ}$ and $f_{7}$ at \SI{480}{G} (solid line) and \SI{10}{G} (dashed line). For $f_{7}$, we fit the theoretical model to the experimental data in order to obtain a more precise value of $A_{\perp}$ for $^{15}\mathrm{NV}$, which we measure to be \SI{3.68(2)}{MHz} at \SI{297}{K}.

The frequency shifts in $f_{DQ}$ and $f_{7}$ due to magnetic field misalignment can be approximated using perturbation theory; see Appendix~\ref{sec:Appendix:PerturbationTheory}. For $f_{DQ}$ the dominant term is a second-order correction, whose fractional shift is described by the following expression:
\begin{equation}
\begin{gathered}
\frac{\Delta f_{DQ}}{2 \cdot {}^{14}\gamma_n B_z} \approx \frac{1}{2} \beta \, \theta^2 \\
\beta = - \frac{\gamma_e}{^{14}\gamma_n} \cdot \frac{4 \left|A_{||}\right| D \, \gamma_e^2 B_z^2}{\left(D^2 - \gamma_e^2 B_z^2\right)^2} \,,
\end{gathered}
\label{eq:Angle14}
\end{equation}
where $\theta$ is the angle between the NV axis and magnetic field, and
$\beta$
$\approx -9.9$ at $B_z = \SI{480}{G}$, and
$\beta$
$\approx \num{-0.003}$ at $B_z = \SI{10}{G}$.

While the second-order correction is similar to that of $f_{DQ}$, for $f_7$ the fourth-order correction is much larger and is described by the following expression:
\begin{equation}
\begin{gathered}
\frac{\Delta f_7}{^{15}\gamma_n B_z} \approx \frac{1}{2} \beta \, \theta^2  \\
\beta = \frac{\gamma_e^2}{^{15}\gamma_n^2} \cdot \frac{4 A_{\perp}^2 D^2}{\left(D^2 - \gamma_e^2 B_z^2\right)^2} \,,
\end{gathered}
\label{eq:Angle15}
\end{equation}
where
$\beta$
$\approx 460$ at $B_z = \SI{480}{G}$, and
$\beta$
$\approx 280$ at $B_z = \SI{10}{G}$.

\subsection{Anisotropy of the hyperfine coupling}
\begin{figure}
\centering
    \includegraphics[width=1\columnwidth]{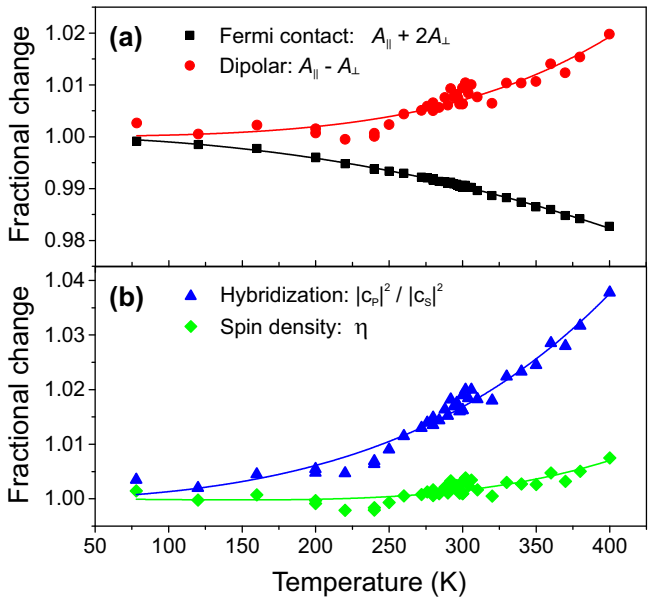}
    \caption{\label{fig:FermiDipolar}
    \textbf{Anisotropy of the $^{14}\mathrm{NV}$ magnetic hyperfine coupling.}
    Temperature dependence of the \textbf{(a)} Fermi contact $(f = A_{||} + 2 A_{\perp})$ and dipolar $(d = A_{||} - A_{\perp})$ terms, \textbf{(b)} NV orbital hybridization $|c_p|^2/|c_s|^2$ and spin density $\eta$ obtained from Eqs.~(\ref{eq:FermiDipolar},\,\ref{eq:SpinHybrid}).
    Markers -- experimental data, solid lines --  polynomial fit.
   }
\end{figure}

The temperature dependence of the magnetic hyperfine coupling components, $A_{||}$ and $A_{\perp}$, is used to obtain the temperature dependence of the Fermi contact and dipolar terms.
These in turn can be expressed in terms of the effective spin density $\eta$ occupying the atomic orbitals of the nitrogen atom and their effective hybridization ratio $|c_p|^2/|c_s|^2$ via the expressions
\begin{alignat}{2} 
    \label{eq:FermiDipolar}
    f &= \frac{8\pi}{3} \frac{\mu_0}{4\pi} g_e \mu_B \mu_n |c_s|^2\eta \, \left|\psi_s(0)\right|^2 \nonumber \\
    &= \SI{1811}{MHz} \times \left(1-|c_s|^2\right)\eta\\
    \label{eq:SpinHybrid} \nonumber
    d &= \frac{2}{5} \frac{\mu_0}{4\pi} g_e \mu_B \mu_n |c_p|^2\eta \,  \left<\psi_p\left|\frac{1}{r^3}\right|\psi_p\right> \nonumber \\
    &= \SI{55.52}{MHz} \times |c_p|^2\eta \, .
\end{alignat}
Figure~\ref{fig:FermiDipolar} shows that both the spin density and hybridization ratio are observed to increase with temperature. This is consistent with the \textit{ab initio} calculations \cite{BAR2019}, which concluded that with increasing temperature, the spin density diffuses away from the three carbon atoms surrounding the vacancy and the nitrogen atom moves towards the vacancy (away from its nearest-neighbor carbon atoms). Thus, one would expect the spin density to increase at the nitrogen atom as a result of the outward diffusion, and the displacement of nitrogen atom would lead to an increase in the hybridization ratio (as the orbitals connecting the nitrogen to its nearest-neighbours must become more p-like to achieve the new geometry of the bond).

\clearpage

\section{\label{sec:ConclusionOutlook}Conclusion and outlook}

We measured the nuclear-spin and electron-spin transition frequencies for NV centers containing $^{14}\mathrm{N}$ and $^{15}\mathrm{N}$, as a function of temperature.
To describe the results, we used numerical diagonalization of the Hamiltonian, including the effect of magnetic field misalignment.
The model allows us to extract the underlying parameters $Q$ (for $^{14}\mathrm{NV}$), $D$, $A_{||}$, $A_{\perp}$, $\gamma_e/\gamma_n$ for both isotopes and their temperature dependences (except $A_{\perp}$ for $^{15}\mathrm{NV}$).
The magnitude of each one of these parameters ($Q$, $D$, $A_{||}$, $A_{\perp}$) decreases with temperature in the range from 77 to 400\,K, showing a reduction of \SI{\sim 0.1}{\percent} (in the case of $Q$) to \SI{\sim 2}{\percent} (in the case of $A_{||}$).

Comparison of the determined parameters reveals a difference in $D$ of \SI{\sim 120}{kHz} (\SI{\sim 40}{ppm}) between NV centers containing $^{14}\mathrm{N}$ and  $^{15}\mathrm{N}$. To our knowledge, this is the first report of such an isotopic difference for NV centers.
We also observe a difference of \SI{\sim 0.1}{\percent} between $^{15}A_{||}/^{14}A_{||}$ and $^{15}\gamma_n/^{14}\gamma_n$.

The temperature dependence of the anisotropy of the hyperfine coupling between electron and nuclear spins ($A_{||}$, $A_{\perp}$) in $^{14}\mathrm{NV}$ can be used to infer the temperature dependence of the Fermi-contact and dipolar interactions, which, in turn, can provide information about the electron spin density and orbital hybridization.

We determined the temperature dependence of $f_{DQ}$ (\SI{+0.52(1)}{ppm/K}) and $f_7$ (\SI{-1.1(1)}{ppm/K}), which are three orders of magnitude smaller than the other nuclear-spin transition frequencies.
Nevertheless, this sensitivity to temperature may limit the performance of nuclear-spin-based sensors and should be taken into account.

We found that residual transverse fields should be carefully considered in order to precisely determine the frequencies, especially for $f_7$, for which a misalignment of the field by 0.1$^{\circ}$ leads to a fractional change of \SI{\sim 600}{ppm}. This strong dependence allowed us to measure $A_{\perp}$ for $^{15}\mathrm{NV}$ to be \SI{3.68+-0.02}{MHz}, which is in agreement with the previously measured value \cite{FEL2009}.

The combination $f_3 - f_6$ has negligible dependence on $A_{\perp}$ and is described to high precision by $f_3 - f_6 = 2 \gamma_{n} B$, and therefore, its temperature dependence is predicted to be weak: $<\SI{10}{ppb \per K}$.

In summary, we have precisely determined the set of parameters relevant for the development of NV-diamond rotation sensors, magnetometers, frequency standards and multisensors, along with the temperature dependence of these parameters. The results indicate a promising path to developing such devices with greatly reduced sensitivity to environmental variations. The general idea is to use multiple transitions with different sensitivity to, for example, temperature, which allows one to isolate the environmental parameter drift from the effect of interest (e.g., inertial rotation).

\textbf{Acknowledgements:} The authors are grateful to Junichi Isoya, Pauli Kehayias and Janis Smits for helpful discussions. S.L. and A.J. acknowledges support from the U.S. Army Research Laboratory under Cooperative Agreement No. W911NF-21-2-0030 and No. W911NF-18-2-0037. V.M.A. acknowledges support from NSF award No. CHE-1945148, NIH awards No. 1R41GM145129 and No. 1DP2GM140921.
This work was supported in part by the EU FET-OPEN Flagship Project ASTERIQS (action 820394) and by the European Commission’s Horizon Europe Framework Program under the Research and Innovation Action MUQUABIS GA no. 101070546.

\appendix

\section{Numerical methods -- Eigenvalues and least square fitting}
\label{sec:Appendix:Numerical}

\begin{figure*}[!t]
\centering
    \includegraphics[width=0.9\columnwidth]{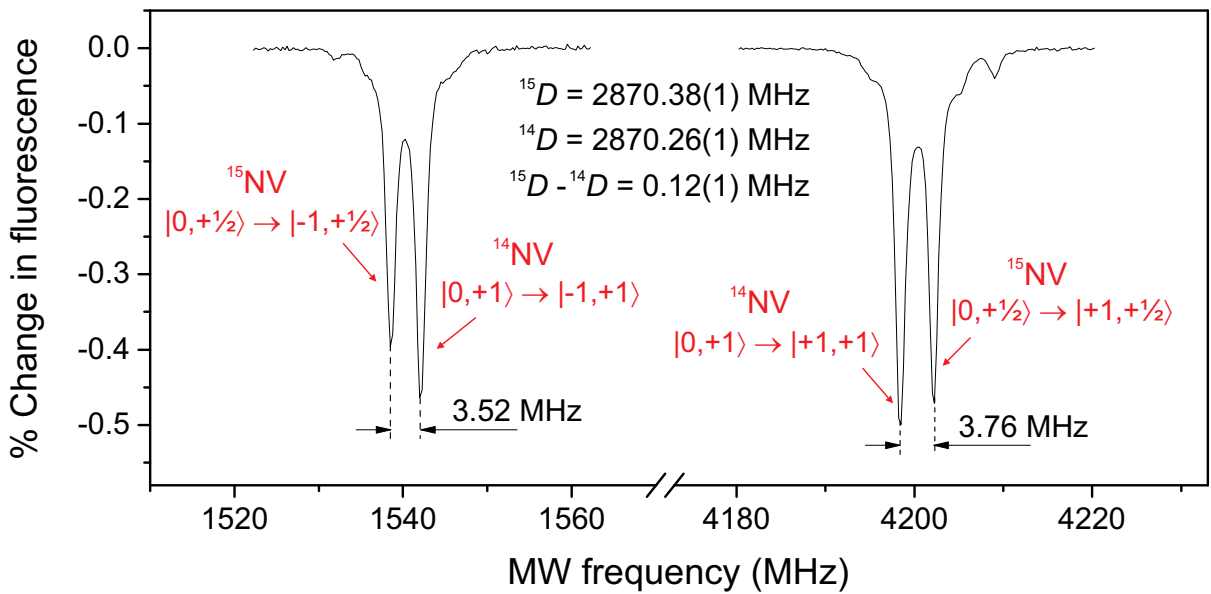}
    \caption{\label{fig:ODMR} \textbf{Measurement of isotopic shift in \textit{D} using ODMR.} The spectrum is obtained from the G3 sample ($50\!:\!50$ isotopic ratio) at \SI{475}{G} and consists of four resonances corresponding to two transitions ($f_+$, $f_-$) for $^{14}\mathrm{NV}$ and two transitions ($f_+$, $f_-$) for $^{15}\mathrm{NV}$. At this field, the nuclear spins are optically polarized to their largest $m_I$ sublevels, $m_I = +1$ for $^{14}\mathrm{NV}$ and $m_I = +1/2$ for $^{15}\mathrm{NV}$, which creates a resolvable splitting.
    A slight difference is observed between the splitting of $f_+$ (\SI{3.76}{MHz}) and that of $f_-$ (\SI{3.52}{MHz}), which corresponds to a difference in $D$ of \SI{0.12 +- 0.01}{MHz}.
   }
\end{figure*}

    The full ground state Hamiltonian of $^{14}\mathrm{NV}$ (and $^{15}\mathrm{NV}$, which is noted in parenthesis throughout this section) is 
    \begin{equation}
    \begin{split}
        H &= D S_z^2 + Q I_z^2 + A_{||} S_z I_z + \gamma_e B_z S_z - \gamma_n B_z I_z \\
        &+ \frac{A_{\perp}}{2} \left(S_+ I_- + S_- I_+\right) + \gamma_e B_x S_x - \gamma_n B_x I_x \label{eq:HamiltonianFull},
    \end{split}
    \end{equation}
    and includes a transverse magnetic field, which is assumed without loss of generality to point in the $+x$ direction. (For $^{15}\mathrm{NV}$, use $Q$ = 0.)
    Given a set of values for the coupling parameters of $^{14}\mathrm{NV}$ (and $^{15}\mathrm{NV}$), which we will denote as the vector \textbf{a} = \Big\langle $D$,\:\allowbreak$\gamma_e B_z$,\:\allowbreak$Q$,\:\allowbreak$A_{||}$,\:\allowbreak$A_{\perp}$,\:\allowbreak$\gamma_e B_x$,\:\allowbreak$\gamma_e / {}^{14}\gamma_n$,\:\allowbreak\bigg($D$,\:\allowbreak$A_{||}$,\:\allowbreak$A_{\perp}$,\:\allowbreak$\gamma_e /{}^{15}\gamma_n$\bigg)\Big\rangle,
    we can use eigenvalue decomposition to numerically obtain the nine (plus six) energy eigenvalues and therefore the frequencies of the six (plus three) RF transitions and two (plus two) MW transitions of the ground state Hamiltonian of the $^{14}\mathrm{NV}$ center, denoted as 
    \textbf{f}(\textbf{a}) = \Big\langle $f_1$,\:\allowbreak$f_2$,\:\allowbreak$f_3$,\:\allowbreak$f_4$,\:\allowbreak$f_5$,\:\allowbreak$f_6$,\:\allowbreak$f_+^{(+1)}$,\:\allowbreak$f_-^{(+1)}$,\:\allowbreak
    \bigg($f_7$,\:\allowbreak$f_8$,\:\allowbreak$f_9$,\:\allowbreak$f_+^{(+1/2)}$,\:\allowbreak$f_-^{(+1/2)}$\bigg)\Big\rangle.
    
    Here, we need to solve the inverse problem, starting with experimental values for the transition frequencies, $\mathbf{f}$, and ending with values for the coupling parameters, $\mathbf{a}$.
    This is done starting with an initial guess for the coupling parameters, $\mathbf{a_0}$, calculating the transition frequencies associated with this guess $\mathbf{f}(\mathbf{a_0})$, and an associated weighted error between the calculated and measured transition frequencies
    $\mathcal{S}(\mathbf{a}) = \sum_{i} \left|\frac{1}{\delta f_i}(f_i(\mathbf{a}) - f_i)\right|^2$,
    where $\delta \bf{f}$ is the measurement uncertainty associated with $\mathbf{f}$.
    The fitted values for the coupling parameters correspond to the value of $\mathbf{a}$ that minimizes $\mathcal{S}(\mathbf{a})$, which was obtained using MATLAB's \textit{fminsearch} function (Nelder-Mead simplex algorithm).
    This process was repeated for each temperature to obtain the temperature dependence of each parameter.
    
    Using the fitted temperature dependence of the coupling parameters, we numerically diagonalize the Hamiltonian at and near \SI{297}{K} for $B = \SI{470}{G}$. This allows us to estimate the temperature dependence of the magnetically sensitive nuclear transitions ($f_1$ to $f_9$), which are shown in Table~\ref{table:297Transitions}. 

\setlength{\tabcolsep}{1pt}
\renewcommand{\arraystretch}{1.2}

\begin{table}
\begin{ruledtabular}
\begin{tabular}{cz{4,5}z{4,6}}
Parameter   &
\multicolumn{1}{c}{\textrm{Value}}  &
\multicolumn{1}{c}{\textrm{Derivative}} \\
&
\multicolumn{1}{c}{\textrm{$\si{kHz}$}}   &
\multicolumn{1}{c}{\textrm{$\si{Hz/K}$}} \\
\colrule

$f_1$       & 5085.95\rlap{$(1)$}  &  -35.2\rlap{$(2)$} \\
$f_2$       & 4799.65\rlap{$(1)$}  &  -35.3\rlap{$(2)$} \\
$f_3$       & 2925.22\rlap{$(8)$}  &  161.5\rlap{$(7)$} \\
$f_4$       & 6970.98\rlap{$(8)$}  & -232.8\rlap{$(7)$} \\
$f_5$       & 7257.28\rlap{$(8)$}  & -232.7\rlap{$(7)$} \\
$f_6$       & 2636.14\rlap{$(8)$}  &  161.5\rlap{$(7)$} \\
\colrule
$f_1-f_2$   & 286.299\rlap{$(2)$}  &  0.149\rlap{$(8)$} \\
$f_5-f_4$   & 286.299\rlap{$(2)$}  &  0.149\rlap{$(8)$} \\
$f_3-f_6$   & 289.081\rlap{$(2)$}  & -0.000\rlap{$(0)$} \\
\colrule
$f_7$       & 205.89\rlap{$(3)$}  & -0.31\rlap{$(2)$}  \\
$f_8$       & 2825.8\rlap{$(1)$}  & -268\rlap{$(2)$}\\
$f_9$       & 3234.8\rlap{$(1)$}  & -269\rlap{$(2)$}\\
\end{tabular}
\caption{\textbf{Transition frequencies and temperature derivatives at $T = \SI{297}{K}$ and $B = \SI{470}{G}$.} Values are determined by performing numeric diagonalization of the Hamiltonian (Eq.\,\ref{eq:Hamiltonian}) using values and uncertainties of $D$, $Q$, $A_{||}$, and $A_{\perp}$ listed in Table~\ref{table:297Parameters}. For $^{15}\mathrm{NV}$, it is assumed that $A_{\perp}$ has the same fractional temperature dependence as $A_{||}$ within \SI{5}{\percent}, or $(\mathrm{d}A_{\perp}/\mathrm{d}T)/A_{\perp} = \num{1.00 +- 0.05} \times (\mathrm{d}A_{||}/\mathrm{d}T)/A_{||}$.}
\label{table:297Transitions}
\end{ruledtabular}
\end{table}

\section{Isotopic shift in $D$} 
\label{sec:Appendix:ODMR}

The isotopic shift in $D$ can be measured directly from the pulsed ODMR spectrum, which is shown in Fig.~\ref{fig:ODMR}.
Signals were recorded for the G3 sample at \SI{475}{G} and \SI{298}{K}. At this field, the nuclear spins are optically polarized to their largest $m_I$ sublevels, $m_I = +1$ for  $^{14}\mathrm{NV}$ and $m_I = +1/2$ for $^{15}\mathrm{NV}$, due to the opposite signs of $^{14}\gamma_n$ and $^{15}\gamma_n$, which allows them to be individually resolved.
By extracting $D$ using $(f_+ + f_-)/2$ for each isotope, we obtain $D = \SI{2870.26}{MHz}$ for $^{14}\mathrm{NV}$ and $D = \SI{2870.38}{MHz}$ for $^{15}\mathrm{NV}$, which corresponds to an isotopic shift in the $D$ parameter of \SI{0.12+-0.01}{MHz}.

\section{Perturbation Theory}
\label{sec:Appendix:PerturbationTheory}

    Perturbation theory can be used to describe how both the transverse hyperfine coupling parameter ($A_{\perp}$) and transverse magnetic fields ($B_x$) shift the nuclear-spin transition frequencies for both $^{14}\mathrm{NV}$ and $^{15}\mathrm{NV}$. Terms in the Hamiltonian [Eq.~(\ref{eq:HamiltonianFull})] can be divided into terms that do ($H_{||}$) and do not ($V$) commute with $S_z$ and $I_z$:
    \begin{align}
    H_{||} &= D S_z^2 + Q I_z^2 + A_{||} S_z I_z + \gamma_e B_z S_z - \gamma_n B_z I_z \\
    V &= \frac{A_{\perp}}{2} \left(S_+ I_- + S_- I_+\right) + \frac{\gamma_e B_x}{2} \left(S_+ + S_-\right) \, .
    \end{align}
    Here we have omitted the transverse nuclear-spin Zeeman term, which is small compared the the transverse electron-spin Zeeman term.
    Using $H_{||}$ as the unperturbed Hamiltonian and treating $V$ as a perturbation, the second-order perturbation shift of each unperturbed state is calculated using the following expression
    \begin{align}
    \Delta E_{m_S,m_I}^{(2)} &= \sum_{m_s^{'},m_I^{'}} \frac{\left|\bra{m_s^{'},m_I^{'}} V \ket{m_s,m_I}\right|^2}{E_{m_S^{\phantom{'}},m_I^{\phantom{'}}}^{(0)} - E_{m_S^{'},m_I^{'}}^{(0)}},
    \end{align}
    where the eigenstates of the unperturbed state are denoted as $\ket{m_s,m_I}$, and their energies are described as follows
    \begin{align}
    E_{m_S,m_I}^{(0)} &= m_s^2 D + m_I^2 Q + m_s m_I A_{||} \nonumber \\
    &+ m_s \gamma_e B_z - m_I \gamma_n B_z .
    \end{align}
    For each nuclear-spin transition, an approximate expression for the total energy shift is obtained to second order in $1/F_{\pm}$, where $F_{\pm} = D \pm \gamma_e B$.
    There are also fourth-order perturbation shifts that produce effects that are of second order in $1/F_{\pm}$, which appear when $m_s^{''}=m_s$, $m_I^{''} \neq m_I$:
\begin{widetext}

\begin{align}
\label{eq:Perturbation4}
\Delta E_{m_S,m_I}^{(4)} &=
\sum_{m_s^{'''},m_I^{'''}}
\sum_{m_I^{''} \neq m_I}
\sum_{m_s^{'},m_I^{'}}
\frac{
    \bra{m_s,m_I} V \ket{m_s^{'''},m_I^{'''}}
    \bra{m_s^{'''},m_I^{'''}} V \ket{m_s^{},m_I^{''}}
    \bra{m_s^{},m_I^{''}} V \ket{m_s^{'},m_I^{'}}
    \bra{m_s^{'},m_I^{'}} V \ket{m_s,m_I}
}
{
    \left(E_{m_S^{\phantom{'}},m_I^{\phantom{'}}}^{(0)} - E_{m_S^{'''},m_I^{'''}}^{(0)}\right)
    \left(E_{m_S^{\phantom{'}},m_I^{\phantom{'}}}^{(0)} - E_{m_S^{},m_I^{''}}^{(0)}\right)
    \left(E_{m_S^{\phantom{'}},m_I^{\phantom{'}}}^{(0)} - E_{m_S^{'},m_I^{'}}^{(0)}\right)
},
\end{align}    
Combining these shifts gives us expressions for the shifted nuclear-spin transition frequencies, both for $^{14}\mathrm{NV}$ and $^{15}\mathrm{NV}$:
\begin{alignat}{3}
    \label{eq:PerturbationShifts14}
    f_1 &= |Q| \phantom{{}-\left|A_{||}\right|} + {}^{14}\gamma_n B_z  - \frac{A_{\perp}^2}{F_-}
        &{}-{}& A_{\perp}^2\left(\frac{|Q| - \left|A_{||}\right|}{F_-^2} + \frac{2|Q| - \left|A_{||}\right|}{F_+^2} \right)
            &{}+{}& \frac{\gamma_e^2 B_x^2}{2} \left[A_{\perp}^2 \frac{3}{Q} \left(\frac{1}{F_+} + \frac{1}{F_-}\right)^2 - \left|A_{||}\right|\left(\frac{1}{F_-^2} - \frac{1}{F_+^2}\right)
                \right] \nonumber \\
    f_2 &= |Q| \phantom{{}-\left|A_{||}\right|} - {}^{14}\gamma_n B_z  - \frac{A_{\perp}^2}{F_+}
        &{}-{}& A_{\perp}^2\left(\frac{2|Q| - \left|A_{||}\right|}{F_-^2} + \frac{|Q| - \left|A_{||}\right|}{F_+^2} \right)
            &{}+{}& \frac{\gamma_e^2 B_x^2}{2} \left[A_{\perp}^2 \frac{3}{Q} \left(\frac{1}{F_+} + \frac{1}{F_-}\right)^2 + \left|A_{||}\right|\left(\frac{1}{F_-^2} - \frac{1}{F_+^2}\right)
                \right] \nonumber \\   
    f_3 &= |Q| - \left|A_{||}\right| + {}^{14}\gamma_n B_z   \phantom{{}- \frac{A_{\perp}^2}{F_+}}
        &{}-{}& A_{\perp}^2\left(\frac{2|Q| - \left|A_{||}\right|}{F_-^2} \right)
            &{}+{}& \frac{\gamma_e^2 B_x^2}{2} \left[A_{\perp}^2 \left(\frac{2}{Q-A_{||}} + \frac{1}{Q+A_{||}}\right)  \frac{1}{F_-^2} + \frac{\left|A_{||}\right|}{F_-^2} \right] \nonumber \\
    f_4 &= |Q| + \left|A_{||}\right| - {}^{14}\gamma_n B_z   + \frac{A_{\perp}^2}{F_-}
        &{}-{}& A_{\perp}^2\left(\frac{|Q|}{F_-^2} \right)
            &{}+{}& \frac{\gamma_e^2 B_x^2}{2} \left[A_{\perp}^2 \left(\frac{1}{Q-A_{||}} + \frac{2}{Q+A_{||}}\right)  \frac{1}{F_-^2} - \frac{\left|A_{||}\right|}{F_-^2} \right] \nonumber \\
    f_5 &= |Q| + \left|A_{||}\right| + {}^{14}\gamma_n B_z   + \frac{A_{\perp}^2}{F_+}
        &{}-{}& A_{\perp}^2\left(\frac{|Q|}{F_+^2} \right)
            &{}+{}& \frac{\gamma_e^2 B_x^2}{2} \left[A_{\perp}^2 \left(\frac{1}{Q-A_{||}} + \frac{2}{Q+A_{||}}\right)  \frac{1}{F_+^2} - \frac{\left|A_{||}\right|}{F_+^2} \right] \nonumber \\
    f_6 &= |Q| - \left|A_{||}\right| - {}^{14}\gamma_n B_z   \phantom{{}- \frac{A_{\perp}^2}{F_+}}
        &{}-{}& A_{\perp}^2\left(\frac{2|Q| - \left|A_{||}\right|}{F_+^2} \right)
            &{}+{}& \frac{\gamma_e^2 B_x^2}{2} \left[A_{\perp}^2 \left(\frac{2}{Q-A_{||}} + \frac{1}{Q+A_{||}}\right)  \frac{1}{F_+^2} + \frac{\left|A_{||}\right|}{F_+^2} \right]
\end{alignat}
\begin{alignat}{6}
    \label{eq:PerturbationShifts15}
    f_7 &=
        &{}\phantom{-}{}&\left|{}^{15}\gamma_n\right| B_z
        &{}+{}& \frac{A_{\perp}^2}{2} \left(\frac{1}{F_-}\right.
            - \left.\frac{1}{F_+}\right)
        &{}+{}&\frac{A_{\perp}^2}{4}\left(
            \frac{A_{||}}{F_-^2} - \frac{A_{||}}{F_+^2}
        \right)
        &{}+{}& \frac{\gamma_e^2 B_x^2}{2}
            \left[\frac{A_{\perp}^2}{\left|{}^{15}\gamma_n\right| B_z} \left(\frac{1}{F_+} + \frac{1}{F_-}\right)^2 - A_{||} \left(\frac{1}{F_-^2} - \frac{1}{F_+^2}\right)\right]&\nonumber \\
    f_8 &= A_{||}
        &{}-{}&\left|{}^{15}\gamma_n\right| B_z
        &{}-{}&\frac{A_{\perp}^2}{2} \left(\frac{1}{F_-}\right)
        &{}-{}&\frac{A_{\perp}^2}{4}\left(
            \frac{A_{||}}{F_-^2}
        \right)
        &{}+{}& \frac{\gamma_e^2 B_x^2}{2}
            \left[\left(\frac{A_{\perp}^2}{A_{||}} - A_{||} \right) \frac{1}{F_-^2}\right] &\nonumber \\
    f_9 &= A_{||}
        &{}+{}& \left|{}^{15}\gamma_n\right| B_z
        &{}-{}&\frac{A_{\perp}^2}{2} \left(\frac{1}{F_+}\right)
        &{}-{}&\frac{A_{\perp}^2}{4}\left(
            \frac{A_{||}}{F_+^2}
        \right)
        &{}+{}& \frac{\gamma_e^2 B_x^2}{2}
            \left[\left(\frac{A_{\perp}^2}{A_{||}} - A_{||} \right) \frac{1}{F_+^2}\right]\,.&
\end{alignat}
\end{widetext}

These expressions can be reduced to obtain simplified expressions for the transition frequencies [see~Eqs.~(\ref{eq:EquationsN14}),(\ref{eq:EquationsN15})], as well as their angular dependences [see~Eqs.~(\ref{eq:Angle14}),(\ref{eq:Angle15})].

For  $^{14}\mathrm{NV}$, we can linearly combine nuclear-spin transition frequencies in order to obtain simple approximations for $^{14}\gamma_n B$, $Q$, and $A_{||}$
\begin{align}
    ^{14}\gamma_n B &\approx \frac{f_3 - f_6}{2} \nonumber \\
    |Q|-|A_{||}| &\approx \frac{f_3 + f_6}{2} \nonumber \\
    |Q| &\approx \frac{f_1 + f_2 + f_3 + f_4 + f_5 + f_6}{6} \nonumber \\
    |A_{||}| &\approx \frac{f_1 + f_2 - 2f_3 + f_4 + f_5 - 2f_6}{6} .
\end{align}


%

\end{document}